\documentclass{article}
\usepackage{spconf,amsmath,epsfig,amssymb}

\newcommand{\nb}{{\bf n}}
\newcommand{\sbb}{{\bf s}}
\newcommand{\Ab}{{\bf A}}

\newcommand{\xb}{{\bf x}}
\newcommand{\hsb}{{\bf \hat s}}
\newcommand{\Lzero}{$\ell^0$-norm}
\newcommand{\Lone}{$\ell^1$-norm}

\newcommand{\Phib}{{\mbox{\boldmath $\Phi$}}}
\newcommand{\phib}{{\mbox{\boldmath $\varphi$}}}

\ninept

\title{Approximate Sparse Decomposition Based on Smoothed $\ell^0$-Norm}
\twoauthors
   {
    H. Firouzi,
    M. Farivar,
    M. Babaie-Zadeh\sthanks{This work has been partially supported
    by Iran National Science Foundation (INSF)
    under contract number 86/994, and also by ISMO and French
    embassy in Iran in the framework of a Gundi-Shapour
    collaboration program.}
    }
   {Sharif University of Technology\\
   Department Of Electrical Engineering\\
   Tehran, Iran}
   {C. Jutten}
   {GIPSA-LAB,\\ Grenoble,\\ France}

\begin{document}

\maketitle

\begin{abstract}
In this paper, we propose a method to address the problem of
source estimation for Sparse Component Analysis (SCA) in the
presence of additive noise. Our method is a generalization of a
recently proposed method (SL0), which has the advantage of
directly minimizing the \Lzero~ instead of \Lone, while being
very fast. SL0 is based on minimization of the smoothed \Lzero~
subject to $\Ab\sbb = \xb$. In order to better estimate the
source vector for noisy mixtures, we suggest then to remove the
constraint $\Ab\sbb = \xb$, by relaxing exact equality to an
approximation (we call our method Smoothed \Lzero~ Denoising or
SL0DN). The final result can then be obtained by minimization of
a proper linear combination of the smoothed \Lzero ~and a cost
function for the approximation. Experimental results emphasize on
the significant enhancement of the modified method in noisy cases.
\end{abstract}


\section{Introduction}

Blind source separation (BSS) consists of detecting the underlying
source signals within some observed mixtures of them without any
prior information about the sources or the mixing system. Let
$\xb\in\mathbb{R}^{n}$ be the vector of observed mixtures and
$\sbb\in\mathbb{R}^{m}$ denote the vector of unknown source signals.
The mixing equation for the linear instantaneous noisy model will
be:
\begin{equation} \label{eq1}
\xb=\Ab\sbb+\nb
\end{equation}
where \Ab{} is the $n \times m$ unknown mixing matrix and \nb{}
denotes the additive noise vector. The aim of BSS is then to
estimate $\sbb$ from observed data \xb{} without any knowledge of
the mixing matrix, \Ab, or the source signals.

In the determined case, when $n$ $\geq$ $m$, the problem can be
successfully solved using 
Independent Component Analysis (ICA)~\cite{Hyv}. However, in the
underdetermined (or over-complete) cases where fewer observations
than sources are provided, even if $\bf A$ is known, there are
infinitely many solutions to the problem since the number of
unknowns exceeds the number of equations. This ill-posedness
could be resolved by the assumption of `Sparsity', i.e. resulting
in non totally blind source separation problem. A signal is
considered to be sparse when only a few of its samples take
significant values. Thus, among all possible solutions of
(\ref{eq1}) we seek the sparsest one, which has then minimum
number of nonzero components, i.e. minimum \Lzero.


SCA can also be viewed as the problem of representing a signal
$\xb\in\mathbb{R}^{n}$ as a linear combination of $m$ vectors,
called \emph{atoms}~\cite{MallZ93}. The atoms $\{ \phib_{i}
\}_{i=1}^{m}$ collectively form a \emph{dictionary}, $n \times m$
matrix, over which the signal is to be decomposed.
There are special interests in the cases where $m>n$ (refer for
example to~\cite{DonoET06} and the references in it). Again we
have the problem of finding the sparsest solution of the set of
underdetermined linear equations $\xb = \sum_{i=1}^{m} s_{i}
\phib_{i}$ where $\Phib \triangleq [\phib_{1},\dots,\phib_{m}]$ is
the dictionary of $m$ atoms. This problem is also called `atomic
decomposition' and has many potential applications in diverse
fields of science~\cite{DonoET06}.

The general Sparse Component Analysis (SCA) problem consists of
two steps: first estimating the mixing matrix, and then finding
the sparsest source vector, assuming the mixing matrix to be
known. The first step can be accomplished by means of clustering
methods~\cite{GribL06}. In this paper, we focus our attention on
the second step; that is for a given mixing matrix, we wish to
find the solution to the following minimization problem:
\begin{equation} \label{eq2}
\hat{\sbb}= \mathrm{argmin}  \| \sbb \|_{0}\qquad \textrm{subject
to }~ \xb=\Ab\sbb
\end{equation}
where $\| \sbb \|_{0}$ denotes the number of non-zero elements of
$\sbb$ (and is usually called the \Lzero ~of $\sbb$).

So far, several algorithms such as Basis Pursuit (BP)
~\cite{ChenDS99,Dono04} and Matching Pursuit (MP)
~\cite{MallZ93,GribL06} have been proposed to approximate the
solution of (\ref{eq2}). The former is based on the observation
that for most large underdetermined systems of linear equations
the minimal \Lone{ } ($\sum_{i}|s_{i}|$ ) solution is also the
sparsest solution~\cite{Dono04}. The minimization of \Lone~can be
efficiently solved using Linear Programming (LP) techniques~\cite{Candes}. 
Despite all recent developments, computational efficiency has still
remained as a main concern.

Recently in ~\cite{Mohimani}, the idea of using \emph{smoothed
$\ell^0$-norm } ~(SL0) was introduced. More precisely this
algorithm minimizes a smooth approximation of the \Lzero ~denoted
by $m-F_{\sigma}(\sbb)$, and the approximation tends to equality
when $\sigma \rightarrow 0$. The algorithm then sequentially
solves the problem:
\begin{equation}
\mathrm{maximize}~ F_{\sigma}(\sbb) \quad\mathrm{s.t.}\quad
\Ab\sbb=\xb
\end{equation}
for a decreasing sequence of $\sigma$.

This approximation accommodates for both continuous optimization
techniques to estimate the sparsest solution of (\ref{eq2}) and a
noise-tolerant algorithm. The idea turned out to be both efficient
and accurate, i.e. providing a better accuracy than \Lone~
minimization algorithms while being about two orders of magnitude
faster~\cite{Mohimani} than LP.

However, the proposed algorithm has not been designed for the noisy
case (\ref{eq1}), where a noise vector, \nb,~has been added to the
observed mixture \xb. In this paper, we will try to generalize the
proposed method to this noisy case by removing the $\Ab\sbb = \xb$
constraint and relaxing the exact equality to an approximation. In
sparse decomposition viewpoint, this means an approximate sparse
decomposition of a signal on an over-complete dictionary. The final
algorithm will then be an iterative minimization of a proper linear
combination of smoothed \Lzero ~and $\|\Ab\sbb - \xb\|_{2}^{2}$.

This paper is organized as follows. Section 2 discusses the main
idea of the proposed method. Section 3 gives a formal statement of
the final algorithm. Finally, experimental results
are presented in Section 4.

\section{Main Idea}
As stated in the previous section, when the dimensions increase,
finding the minimum \Lzero ~solution of (\ref{eq2}) is impractical
for two reasons. Firstly because \Lzero ~of a vector is a
discontinuous function of its elements and leads to an intractable
combinatorial optimization, and secondly because of the solution
being highly sensitive to noise. The idea of~\cite{Mohimani} is then
to replace the \Lzero ~by continuous function, which approximates
Kronecker delta function, and use optimization techniques to
minimize it subject to $\Ab\sbb=\xb$, as a constraint. For example,
consider the Gaussian like function:
\begin{equation} \label{eq3}
  F_\sigma(\sbb)=\sum_{i=1}^m \exp{(-s_i^2/2\sigma^2)}
\end{equation}
where $s_{i}$ denotes the $i$-th element of vector $\sbb$. For
sufficiently small values of $\sigma$, $F_{\sigma} (s )$ tends to
count the number of zero elements of the vector s. Thus we have:
\begin{equation} \label{eq4}
  \|\sbb\|_0=m-\lim_{\sigma \to 0}F_\sigma(\sbb)
\end{equation}
where $m$ is the dimension of the vector $\sbb$. The sparsest
solution of (\ref{eq2}) can then be approximated by the solution of
the following minimization problem:
\begin{equation} \label{eq5}
\hat{\sbb}=\mathrm{argmin}~(m-F_\sigma(\sbb))\qquad
\textrm{subject to }~ \xb=\Ab\sbb
\end{equation}
The above minimization task can be accomplished using common
gradient type (e.g. steepest descent) algorithms. Note that the
value of $\sigma$ determines how smooth the function $F_\sigma$
is; the smaller the value of $\sigma$, the better the estimation
of $\|\sbb\|_0$ but the larger the probability of being trapped in
local minima of the cost function. 
The idea of~\cite{Mohimani} for escaping from local minima is
then to use a decreasing set of values for $\sigma$ in each
iteration. More precisely for each value of $\sigma$ the
minimization algorithm is initiated with the minimizer of the
$F_{\sigma}(\sbb)$ for the previous (larger) value of $\sigma$.

Now consider a more realistic case where a noise vector, \nb, has
been added to the observed mixture, as in (\ref{eq1}). Here we
notice that we have an uncertainty on exact value of the observed
vector and it seems reasonable to remove the $\xb=\Ab\sbb$
constraint and reduce it to $\xb\approx \Ab\sbb$. This idea is
based on the observation that in presence of considerable noise,
this constraint may lead to a totally different sparse
decomposition. Thus we wish to minimize two terms;
$\|\Ab\sbb-\xb\|_2$ as cost of approximation, and the smoothed
\Lzero ~($m-F_\sigma (\sbb )$), as the measure of
sparsity.

For the sake of simplicity, we choose $\|\Ab\sbb-\xb\|_2^2$ as the
cost of approximation. Therefore, the idea will naturally leads us
to the following minimization problem:
\begin{equation} \label{eq6}
\hat{\sbb}=\mathrm{argmin}~J_\sigma (\sbb) = (m-F_\sigma(\sbb))+
\lambda~ \|\Ab\sbb-\xb\|_2^2
\end{equation}
where $\lambda>0$, represents a compromise between the two terms
of our cost function; sparsity and equality condition.
Intuitively, we may expect that for less noisy mixtures, the value
of $\lambda$ should be greater than that of observations with high
noise quantity.
Further discussion on the choice of $\lambda$ is left to Section
\ref{sec:exp results}.

Another advantage of removing $\xb=\Ab\sbb$ constraint appears
when the dictionary matrix, $\Ab$, is not full rank. In this case
satisfying the exact equality constraint for observed vectors,
which are not in column space of $\bf A$ is impossible and as a
result the previous algorithm fails to find any answer.

\section{Final Algorithm}
\label{sec:FinalAlgorithm} The final algorithm is shown in
Fig~\ref{fig:algorithm}. We call our algorithm SL0 DeNoising
(SL0DN). As seen in the algorithm, the final values of the
previous estimation are used for the initialization of the next
steepest descent step. The decreasing sequence of $\sigma$ is
used to escape from getting trapped into local minima.

Direct calculations show that: 
\begin{eqnarray}
\Delta\sbb=\frac{\partial J_{\sigma}(\sbb)}{\partial \sbb}= \lambda(2\Ab^T(\Ab\sbb-\xb)) \nonumber \\
 +\frac{1}{\sigma^2}[s_{1}e^{(-s_{1}^{2}/2\sigma^{2})},
\dots,s_{m}e^{(-s_{m}^{2}/2\sigma^{2})}]^{T}
\end{eqnarray}

In the minimization part, the steepest descent with variable
step-size ($\mu$) has been applied: If $\mu$ is such that
$J_{\sigma}(\sbb-\mu\Delta{\sbb})<J_{\sigma}(\sbb)$ we multiply it
by 1.2 for the next iteration, otherwise it is multiplied by 0.5.

\begin{figure}[htb]

\vrule
\begin{minipage}{8.5cm} 
\hrule \vspace{0.5em} 
\begin{minipage}{7.5cm} 
\begin{itemize}
        \item Initialization:

           \begin{enumerate}
             \item Let $\hsb_0=\Ab^{T}(\Ab\Ab^{T})^{-1}\xb$.

             \item Choose a suitable value for $\lambda$ as a function of
             $\sigma_{\mathrm{n}}
             $. The value of $\sigma_{\mathrm{n}}$ for a set of observed mixtures may be estimated 
             either directly from the observed mixtures (see for example~\cite{Zayyani} and references therein)
             or using a bootstrap method (discussed in experiment 1 of Section~\ref{sec:exp
             results}).

             \item Choose a suitable decreasing sequence for
             $\sigma$,\\
             $[\sigma_{1}\ldots\sigma_{K}]$. and a sufficiently
             small value for the step-size parameter, $\mu$.
           \end{enumerate}

        \item For $k=1,\dots,K$:
          \begin{enumerate}
             \item Let $\sigma=\sigma_k$.
             \item Minimize (approximately) the function
             $J_{\sigma}(\sbb)$ using $L$ iterations of the
             steepest descent algorithm:
             \begin{itemize}
               \item Initialization: $\sbb\leftarrow\hsb_{k-1}$.
               \item for $j=1\dots L$ (loop $L$ times):
                    \begin{enumerate}
                       \item Let:
                       $\Delta\sbb=\lambda(2\Ab^T(\Ab\sbb-\xb))+\frac{1}{\sigma^2}[s_{1}e^{(-s_{1}^{2}/2\sigma^{2})},
                       \dots,s_{m}e^{(-s_{m}^{2}/2\sigma^{2})}]^{T}$


                       \item If $J_{\sigma}(\sbb-\mu\Delta{\sbb})<J_{\sigma}(\sbb)$
                       let $\rho=1.2$ else $\rho=0.5$.

                       \item Let $\sbb\leftarrow \sbb-\mu \Delta\sbb$ 


                       \item Let $\mu\leftarrow \mu\times\rho$. (variable step-size)
                       \item Set $\hsb_{k}\leftarrow\sbb$.
                    \end{enumerate}
             \end{itemize}

          \end{enumerate}
        \item Final answer is $\hsb=\hsb_K$.
\end{itemize}

\end{minipage}
\vspace{0.5em} \hrule
\end{minipage}\vrule \\
\\
\caption{The final algorithm of SL0DN.} \label{fig:algorithm}
\end{figure}

\section {Experimental Results}
\label {sec:exp results} In this section we investigate the
performance of the proposed method and present our simulation
results. Since our framework is a generalization of the idea
presented in~\cite{Mohimani}, the practical considerations in
that paper can be directly imported into our framework.

In~\cite{Mohimani}, it has been experimentally shown that SL0 is
about two orders of magnitude faster than the state-of-the-art
interior-point LP solvers~\cite{Candes}, while being more
accurate. We provide the comparison results of our method with
the SL0 method. Moreover a comparison with Basis Pursuit
Denoising will be presented.

In all experiments, sparse sources have been artificially
generated using a Bernoulli-Gaussian model: each source is
`active' with probability $p$, and is `inactive' with probability
$1-p$. If it is active, its value is modeled by a zero-mean
Gaussian random variable with variance $\sigma^2_\mathrm{on}$; if
it is not active, its value is modeled by a zero-mean Gaussian
random variable with variance $\sigma^2_\mathrm{off}$, where
$\sigma^2_\mathrm{off} \ll \sigma^2_\mathrm{on}$. Consequently,
each $s_i$ is distributed as:
\begin{equation}
\label{eq: the sources model} s_{i}\sim p \cdot
\mathcal{N}(0,\sigma_{\mathrm{on}})+(1-p) \cdot
\mathcal{N}(0,\sigma_{\mathrm{off}}),
\end{equation}
Sparsity implies that $p \ll 1$. We considered $p=0.1$,
$\sigma_{\mathrm{off}}=0.01$ and $\sigma_{\mathrm{on}}=1$. Elements
of the mixing matrix, \Ab, and noise vector, \nb, were also
considered to have normal distributions with standard deviation of 1
and $\sigma_{\mathrm{n}}$, respectively. As in~\cite{Mohimani}, the
set of decreasing values for $\sigma$ was fixed to
$[1,0.5,0.2,0.1,0.05,0.02,0.01]$.

\vspace{1em} \noindent \textbf{Experiment 1. Optimal value of
$\lambda$}

\begin{figure}[htb]
  \centerline{\epsfig{figure=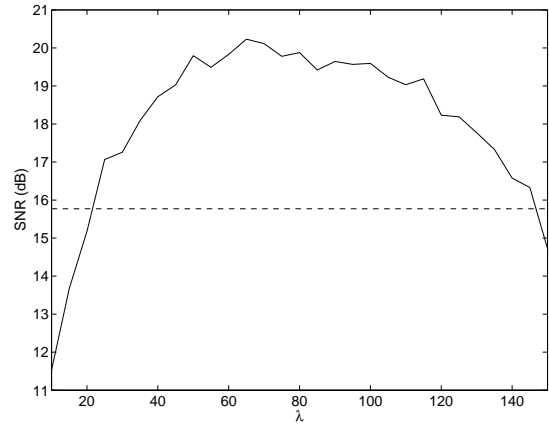,width=8.5cm}}
  \caption{Average Output SNR for different choices of $\lambda$ for $\sigma_{\mathrm{n}}$=0.05. }
\label{fig:SampleExp}
\end{figure}

In this experiment, we investigate the effect of $\lambda$ on the
performance of our method. We set the dimensions to $m=1000$,
$n=400$, and for each value of
$\sigma_{\mathrm{n}}=0,0.01,\dots,0.15$ we plotted the average
Signal to Noise Ratio (SNR), defined by $ 10\log_{10}\frac{\Vert
\sbb \Vert^{2}}{\Vert \hsb-\sbb \Vert^{2}}$, as a function of
$\lambda$ (in this section, all the results are averaged over 100
experiments). Figure~\ref{fig:SampleExp} shows a sample of our
experiments. Dash line represents the results obtained from
(\ref{eq5}), which is independent of $\lambda$. Note that, there
exists an interval in which the choice of $\lambda$ will result
in a better estimation compared to SL0. The SNR takes its maximum
in this region for some value of $\lambda$, which we call
$\lambda_{\mathrm{opt}}$.

\begin{figure}[htb]
  \centerline{\epsfig{figure=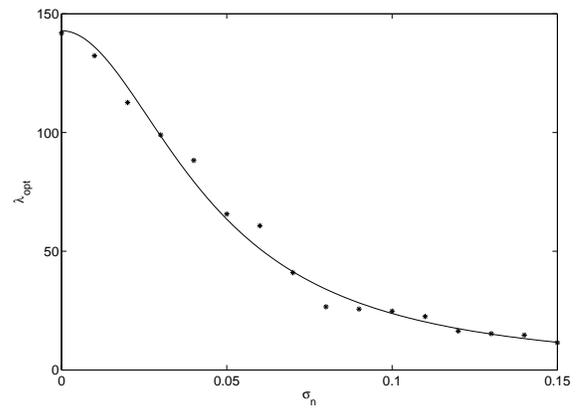,width=8.5cm}}
  \caption{$\lambda_{\mathrm{opt}}$ as a function of noise power ($\sigma_{\mathrm{n}}$). The continuous curve shows our
  approximation of $\lambda_{\mathrm{opt}}$.}
\label{fig:Interpolation}
\end{figure}

As mentioned in the previous section, we expect an appropriate
choice of $\lambda$ to be a decreasing function of
$\sigma_{\mathrm{n}}$ since with the increase of noise power, the
cost of approximation $(\Ab \xb \approx \sbb)$ decreases. To
verify this, 
for each value of $\sigma_{\mathrm{n}}$, we obtained the value of
$\lambda_{\mathrm{opt}}$ using the curves similar to
Fig.~\ref{fig:SampleExp}. Figure~\ref{fig:Interpolation} shows
the values of $\lambda_{\mathrm{opt}}$ as a function of
$\sigma_{\mathrm{n}}$ in [0,0.15]. We fit these results with a
curve of type $\frac{1}{\alpha + \beta x^2}$ to find the
following rule of thumb for the choice of parameter $\lambda$:


\begin{equation} \label{eq7}
\lambda \approx \frac{1}{0.007+3.5 \sigma_{\mathrm{n}}^2}.
\end{equation}
This formula gives a rough approximation for the choice of
appropriate $\lambda$ in the initialization step of the algorithm.

Notice that we have two choices for the initialization of the
proposed method: either to estimate $\sigma_{\mathrm{n}}$
directly from the observed mixtures \cite{Zayyani} and then use
(\ref{eq7}) to find an approximation of $\lambda_{\mathrm{opt}}$,
or to follow this iterative approach to solve the problem:
\begin{enumerate}
\item{choose an arbitrary reasonable value of $\sigma_{\mathrm{n}}$.}
\item{take $\lambda_{\mathrm{opt}}$ from the curve.}
\item{run the algorithm and after convergence, compute an estimation of $\sigma_{\mathrm{n}}$ from the
obtained source vector and then goto step 2.}
\end{enumerate}


\vspace{1em} \noindent \textbf{Experiment 2. Speed and
performance}

In order to measure the speed of our algorithm, we run the
algorithm 100 times for $m=1000$, $n=400$ and
$\sigma_{\mathrm{n}}=0.05$. The simulation is performed in MATLAB7
environment using an Intel 2.8Ghz processor and 512MB of memory.
The average run time of SL0DN was 2.062 seconds while the average
time for SL0 was 0.242 seconds. Although SL0DN is somehow slower
than SL0, but regarding to Table I in~\cite{Mohimani}, the
algorithm is still much faster than $\ell_1$-magic and FOCUSS.

\begin{figure}[tb]
  \centerline{\epsfig{figure=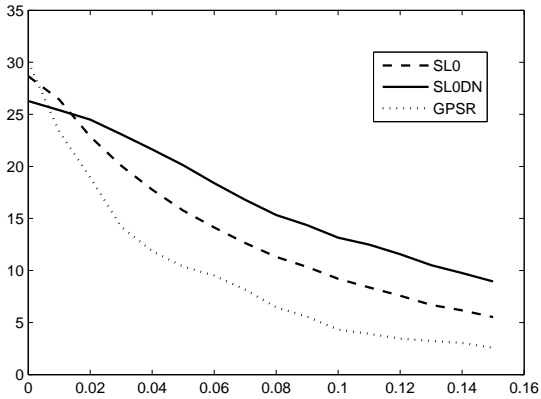,width=8.5cm}}
  \caption{Comparison between SL0DN, SL0 and BPDN.}
\label{fig:compare3}
\end{figure}

We proceed with the performance analysis of the proposed
algorithm. In this experiment, we fix the parameters $m, n, p$
with those of experiment 1 and for each value of
$\sigma_{\mathrm{n}}$, choose the value of $\lambda$ with
(\ref{eq7}). In Fig.~\ref{fig:compare3} the average output SNR is
compared to the results of SL0. It can be seen that except for
low-noise mixtures $(\sigma_{\mathrm{n}}<0.02)$, SL0DN achieves a
better SNR. Thus for noisy mixtures, the case for most real data,
the act of approximately satisfying $\Ab \sbb=\xb$ constraint  is
justified experimentally.


We also compared the results SL0DN with Basis Pursuit DeNoising
(BPDN) which is much faster than BP. We used Gradient Projection
for Sparse Reconstruction (GPSR)~\cite{GPSR} algorithm
for BPDN. The results of GPSR are shown in
Fig.~\ref{fig:compare3} with dotted line. As we see, the average
SNR curve of GPSR lies under the two other curves except for low
noise mixtures. It worths mentioning that 
the average run time of GPSR was 3.156 seconds.

\vspace{1em} \noindent \textbf{Experiment 3. Dimension Dependency}

\begin{figure}[tb]
  \centerline{\epsfig{figure=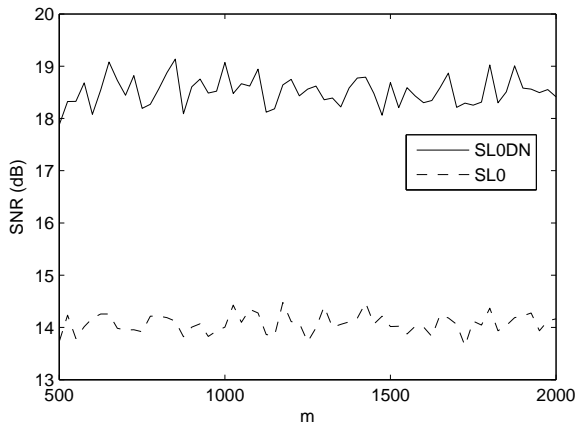,width=8.5cm}}
  \caption{Average Output SNR versus $m$. Averages are taken over $100$ experiments.}
\label{fig:compare2}
\end{figure}

In this experiment we study the performance of the proposed method
for different dimensions of sources and mixtures. In this
experiment, 
the values of $m$ and $n$ change within a constant ratio
($n=0.4m$). The average output SNR for both methods are shown in
Fig.\ref{fig:compare2}. The results suggest that the quality of
estimation is almost independent of the dimensions.

\section{conclusion}
\label{sec:conclusion} We presented a fast method for Sparse
Component Analysis (SCA) or atomic decomposition on over-complete
dictionaries, in presence of additive noise. The method was a
generalization of SL0 method. The proposed method was based on
smoothed \Lzero\\~minimization and satisfying the equality
constraint approximately instead of exact equality constraint.
The proposed method is fast while being more robust against noisy
mixtures than the original SL0. Experimental results approved the
performance and the noise-tolerance of our method for noisy
mixtures.


\begin{thebibliography}{99}

\bibitem{Hyv}
A.~Hyv\"arinen, J.~Karhunen, and E.~Oja,
\newblock {\em Independent Component Analysis},
\newblock John Wiley \& Sons, 2001.


\bibitem{MallZ93}
S.~Mallat and Z.~Zhang,
\newblock ``Matching pursuits with time-frequency dictionaries,''
\newblock {\em IEEE Trans. on Signal Proc.}, vol. 41, no. 12, pp. 3397--3415,
  1993.

\bibitem{DonoET06}
D.~L. Donoho, M.~Elad, and V.~Temlyakov,
\newblock ``Stable recovery of sparse overcomplete representations in the
  presence of noise,''
\newblock {\em IEEE Trans. Info. Theory}, vol. 52, no. 1, pp. 6--18, Jan 2006.

\bibitem{GribL06}
R.~Gribonval and S.~Lesage,
\newblock ``A survey of sparse component analysis for blind source separation:
  principles, perspectives, and new challenges,''
\newblock in {\em Proceedings of ESANN'06}, April 2006, pp. 323--330.



\bibitem{ChenDS99}
S.~S. Chen, D.~L. Donoho, and M.~A. Saunders,
\newblock ``Atomic decomposition by basis pursuit,''
\newblock {\em SIAM Journal on Scientific Computing}, vol. 20, no. 1, pp.
  33--61, 1999.


\bibitem{Dono04}
D.~L. Donoho,
\newblock ``For most large underdetermined systems of linear equations the
  minimal $\ell^{1}$-norm solution is also the sparsest solution,''
\newblock Tech. {R}ep., 2004.

\bibitem{Candes}
E. Candes and J. Romberg, ``$\ell_1$-magic: Recovery of sparse
signals via convex programming'' 2005, URL:
www.acm.caltech.edu/l1magic/downloads/l1magic.pdf.



\bibitem{Mohimani}
G.~H. Mohimani, M.~Babaie-Zadeh, and C.~Jutten,
\newblock ``Fast sparse representation based on smoothed $l^0$-norm,''
\newblock accepted for publication in {\em IEEE Trans. on Signal
Proc.} (available as eprint arXiv 0809.2508).






\bibitem{Zayyani}
H.~Zayyani, M.~Babaie-Zadeh and C.~Jutten ``Source estimation in
noisy Sparse Component Analysis,'' 15'th Intl. Conf. on Digital
Signal Processing (DSP2007), pp.~219-222 July 2007.



\bibitem{GPSR}
Mario~A.T.~Figueiredo, Robert~D.~Nowak, and Stephen~J.~Wright,
``Gradient projection for sparse reconstruction: Application to
compressed sensing and other inverse problems'', IEEE Journal of
Selected Topics in Signal Processing: Special Issue on Convex
Optimization Methods for Signal Processing, 1(4), pp. 586-598,
2007









\end{thebibliography}
\end{document}